\documentstyle[epsfig,12pt,twoside,espcrc1]{article}

\newcommand{\be}{\begin{equation}}
\newcommand{\ee}{\end{equation}}
\newcommand{\bea}{\begin{eqnarray}}
\newcommand{\eea}{\end{eqnarray}}
\newcommand{\bean}{\begin{eqnarray*}}
\newcommand{\eean}{\end{eqnarray*}}

\newcommand{\AmS}{{\protect\the\textfont2
  A\kern-.1667em\lower.5ex\hbox{M}\kern-.125emS}}

\title{Search for Glueballs from Three-body Annihilation of
$\bar pp$ in-Flight }

\author{Bing-Song Zou\address{Queen Mary and
Westfield College, London E1 4NS, UK}\thanks{Now at Institute of High
Energy Physics, P.O.Box 918(4), Beijing 100039, China}}

\begin{document}
\maketitle

\begin{abstract}
Lattice QCD and other theoretical models predict that the $0^{-+}$,
$2^{++}$ and $2^{-+}$ glueballs have masses in the range of 2.0 to 2.4
GeV. For resonances in such an energy range, three-body decay modes are
expected to be large. The strategy for looking for these glueballs
and the newest results from studying Crystal Barrel
data on the three-body annihilation of $\bar pp$ in flight are
presented.
\end{abstract}

\section{STRATEGY FOR GLUEBALL HUNTING IN 
$p\bar p$ ANNIHILATION}

Proton-antiproton annihilation is regarded as a favorable process for glueball 
production. For $p\bar p$ annihilation, there are two possible ways to
produce glueballs as shown in Fig.1(a\&b), which are so called 
``production" and ``formation" mechanisms, respectively.

\vspace*{-0.8cm}
\begin{figure}[htbp]
\vspace*{-0.8cm}
\begin{center}\hspace*{-0.cm}
\epsfysize=6cm
\epsffile{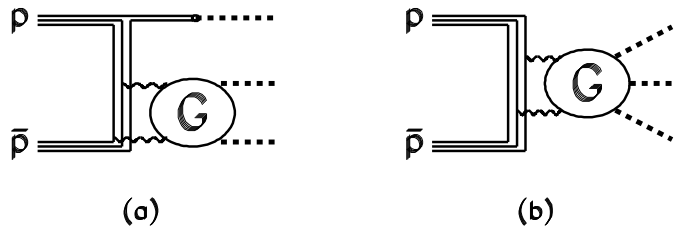}
\end{center}
\vspace*{-2.0cm}
\caption{Glueball production mechanisms for $p\bar p$ annihilation: 
(a) ``production" mechanism; (b) ``formation" mechanism. Dashed and helix lines
are for mesons and gluons, respectively. }
\end{figure}

For the $0^{++}$ glueball ground state, lattice QCD predicts its mass to be
$1.45\sim 1.8 GeV/c^2$ \cite{Lattice}. For a glueball in such a mass range
to be produced from $p\bar p$ annihilation,
it can only come from the ``production" mechanism. 
Indeed, by studying $p\bar p\to 3\pi^0$ \& $\pi^0\eta\eta$, Crystal Barrel
Collaboration\cite{1500}  discovered the $f_0(1500)$ resonance which is now
regarded as the best $0^{++}$ glueball candidate\cite{Braune}. 
There is also some new evidence for $f_0(1770)$\cite{Bugg1}. 

If the $f_0(1500)$ is really a glueball, it
sets a mass scale for glueballs of other quantum numbers.
The $0^{-+}$, $2^{++}$ and $2^{-+}$ glueballs are predicted to be around 
$2.1\sim 2.4$ GeV by various theoretical models\cite{Lattice,QCDsum,Cui}. 
For resonances in such an energy range, they should be mainly produced
from the ``formation" mechanism and three-body decay modes are
expected to be large. 

Crystal Barrel has taken a lot of data for all neutral final states in flight.
Then the question is which three body channels we should study first.
Here we can take some lessons from charmonium decays.
For $\eta_c(0^{-+})$, $\chi_{c0}(0^{++})$, $\chi_{c2}(2^{++})$ 
hadronic decays and $J/\Psi$ radiative decays, they definitively go through 
two-gluon intermediate states. The $\eta$, $\eta'$, $\sigma$ and $f_0(1500)$
seem to be favoured decay products of two-gluon states\cite{PDG,Bugg2}. 
The $\eta\sigma$, $\eta'\sigma$ and $\eta f_0(1500)$ are expected to be
large decay modes of $0^{-+}$ glueballs while the $\eta f_2$ and $\eta'f_2$
are expected to be large for $2^{-+}$ and $2^{++}$ glueball decays.
These decay modes have $\pi^0\pi^0\eta$, $\pi^0\pi^0\eta'$ and $3\eta$
as their final states. 

Hence our strategy for hunting $0^{-+}$,
$2^{++}$ and $2^{-+}$ glueballs is to study resonances formed by
$p\bar p$ and decaying into $\pi^0\pi^0\eta$, $\pi^0\pi^0\eta'$ and $3\eta$
final states first.

\section{ STATUS OF THREE-BODY ANNIHILATION IN-FLIGHT}

Crystal Barrel at LEAR has collected data triggering on neutral final states
at beam momenta 0.6, 0.9, 1.05, 1.2, 1.35, 1.525, 1.642, 1.8 and 1.94 GeV/c,
which correspond to center-of-mass energies ranging from 1.96 to 2.41 
$GeV/c^2$. An average of 8.5 million all neutral events were taken at
each momentum.

These data have been processed. Rough number of selected events at each 
beam momentum and background level for
reconstructed three-body channels from $6\gamma$ and $7\gamma$ events are
listed in Table 1. Among them, the background level for the
$\pi^0\pi^0\omega$
channel has not been investigated yet; the background level for 
$\pi^0\pi^0\eta'$ is too high to do partial wave analysis. So we have been
analyzing the first four channels.

\begin{table}[hbt]
\newlength{\digitwidth} \settowidth{\digitwidth}{\rm 0}
\catcode`?=\active \def?{\kern\digitwidth}
\caption {Rough number of selected events at each beam momentum and
background level for
reconstructed three-body channels from $6\gamma$ and $7\gamma$ events.}
\begin{center}
\begin{tabular}{ccc} 
\hline                                       
channel & Number of Events &  background level  \\\hline
$\pi^0\pi^0\pi^0$ & $\sim 150K$ & 1\% \\
$\pi^0\pi^0\eta$ & $\sim 70K$ & 3\% \\
$\pi^0\eta\eta$ & $\sim 6K$ & 6\% \\
$\eta\eta\eta$ & $\sim 150$ & 3\% \\
$\pi^0\pi^0\eta'$ & $\sim 1K$ & 50\% \\
$\pi^0\pi^0\omega$ & $\sim 70K$ &   \\\hline  
\end {tabular}
\end{center}
\end{table}

For $\pi^0\pi^0\pi^0$ channel, the most obvious contributions come from 
$f_2(1270)\eta$ and $f_0(1500)\eta$ intermediate states. The partial wave
analysis is in progress.

For $\pi^0\eta\eta$ channel, the statistics is not enough for a full
partial analysis including both production and formation amplitudes.
Some effective formalism was used to concentrate on searching for
resonances in the production mechanism\cite{Bugg1}. The main results
are: $f_0(1500)\to\eta\eta$ is clearly seen; $f_0(1750\sim 1800)$,
$f_0(2100)$ and $a_2(1660)$ are confirmed; in addition there is a broad
$f_2(1980)\to\eta\eta$ with mass $M=1980\pm 50$ MeV and width 
$\Gamma=500\pm 100$ MeV.

For $\eta\eta\eta$ channel, the statistics is low. But we can still learn
something. In Fig.2, we show the real data and some Monte Carlo Dalitz 
plots for $\bar pp\to\eta\eta\eta$ at 1.8 GeV/c. 

\vspace*{-0.5cm}
\begin{figure}[htbp]
\vspace*{-0.5cm}
\begin{center}\hspace*{-0.cm}
\epsfysize=14cm
\epsffile{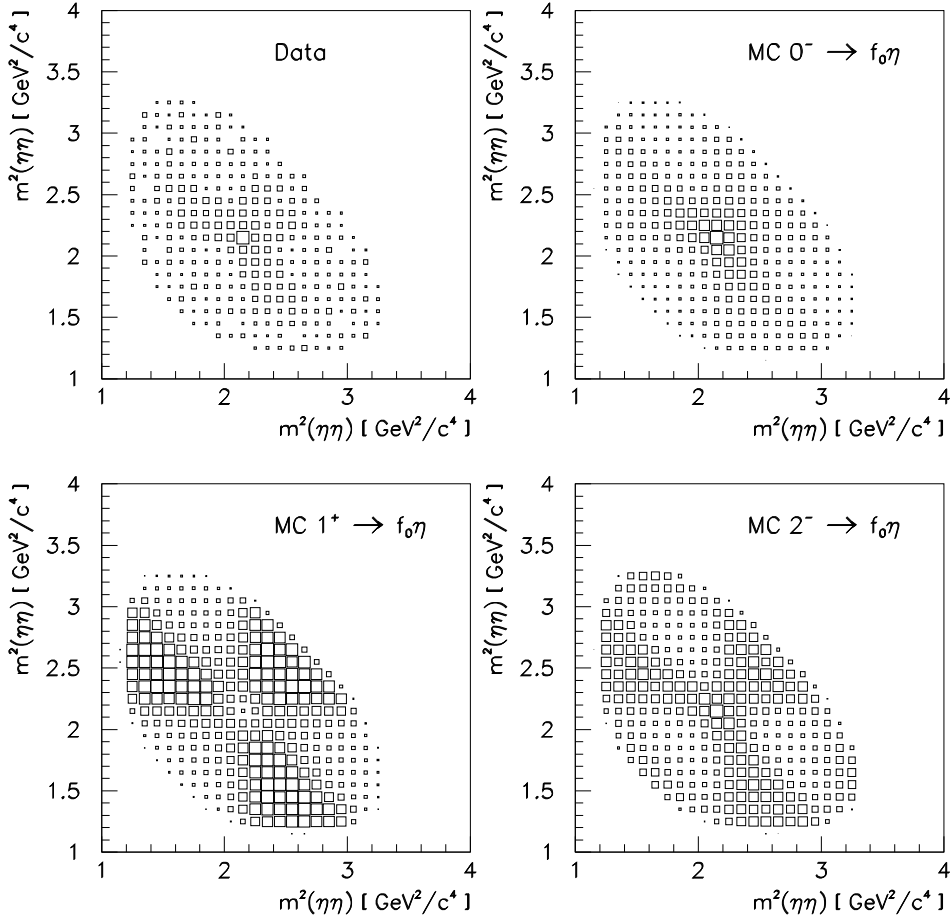}
\end{center}
\vspace*{-1.0cm}
\caption{Real data and Monte Carlo Dalitz plots for 
$p\bar p\to\eta\eta\eta$ at 1.8 GeV/c. }
\end{figure}

From the real data Dalitz plot in Fig.2, the $f_0(1500)$ is obviously
there.
For the beam momentum of 1.8 GeV/c, the orbital angular momentum between
$f_0(1500)$ and $\eta$ is expected to be $\leq 2$, which corresponds to
the initial states of $0^-(l_f=0)$, $1^+(l_f=1)$ and $2^-(l_f=2)$.
From their Monte Carlo Dalitz plots in Fig.2, without fitting data,
it is already clear that the $0^-\to f_0(1500)\eta$ is the most obvious
contribution to $\bar pp\to\eta\eta\eta$.

Preliminary results of the cross section for $\bar pp\to\eta\eta\eta$ is
shown in Fig.3. There are two peaks at about 2.15 GeV and 2.33 GeV.
They may be due to statistical fluctuations. But if they are due to two
resonances, they are most likely $0^{-+}$ resonances with the
$f_0(1500)\eta$ decay mode.

\begin{figure}[htbp]
\begin{minipage}[t]{70mm}
\vspace*{-0.3cm}
\epsfysize=7cm
\epsffile{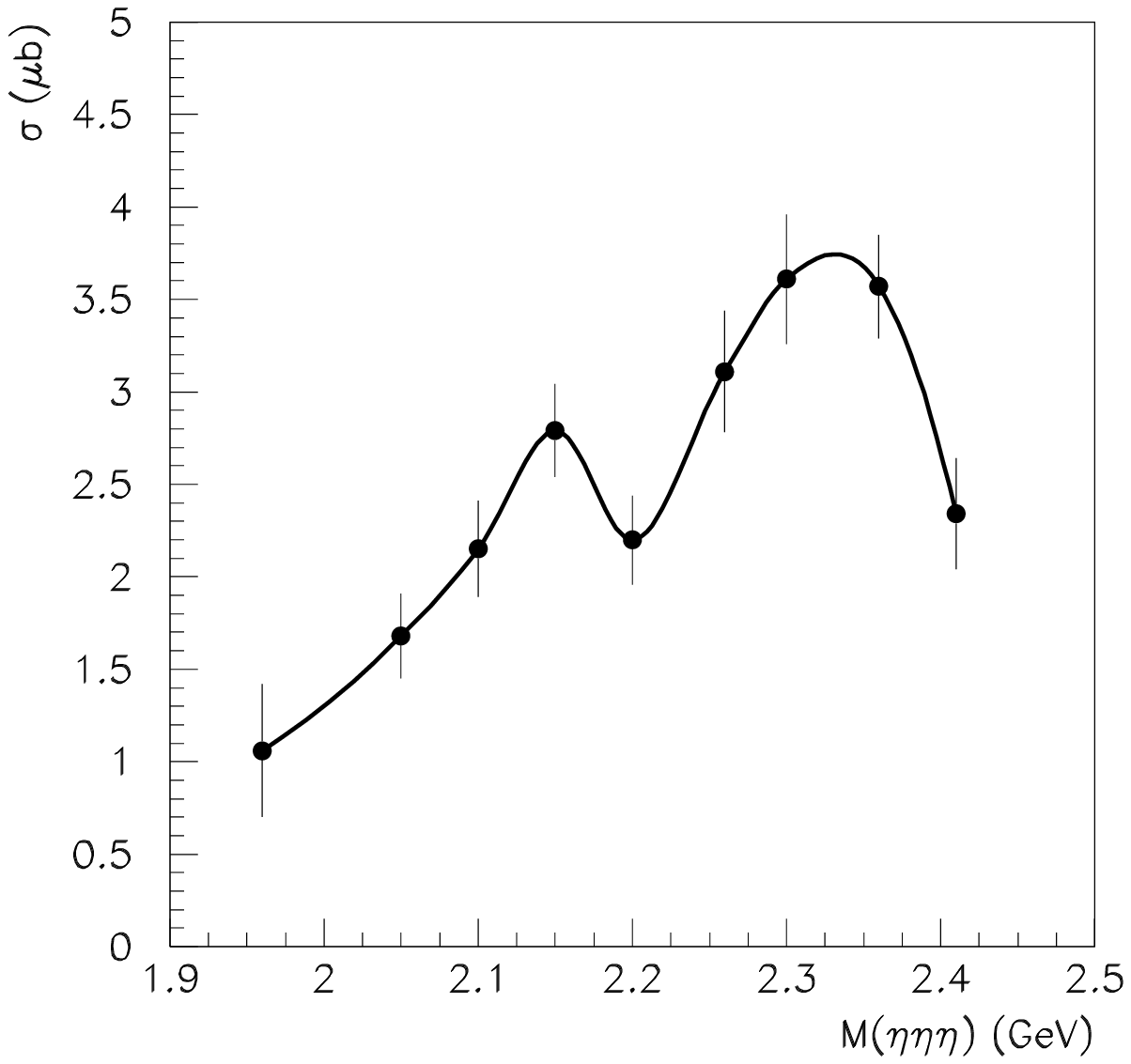}
\vspace*{-1.5cm}
\caption{preliminary results of cross section for $\bar pp\to\eta\eta\eta$. }
\end{minipage}
\hspace*{8.5cm}
\begin{minipage}[t]{70mm}
\vspace*{-8.2cm}
\epsfysize=7cm
\epsffile{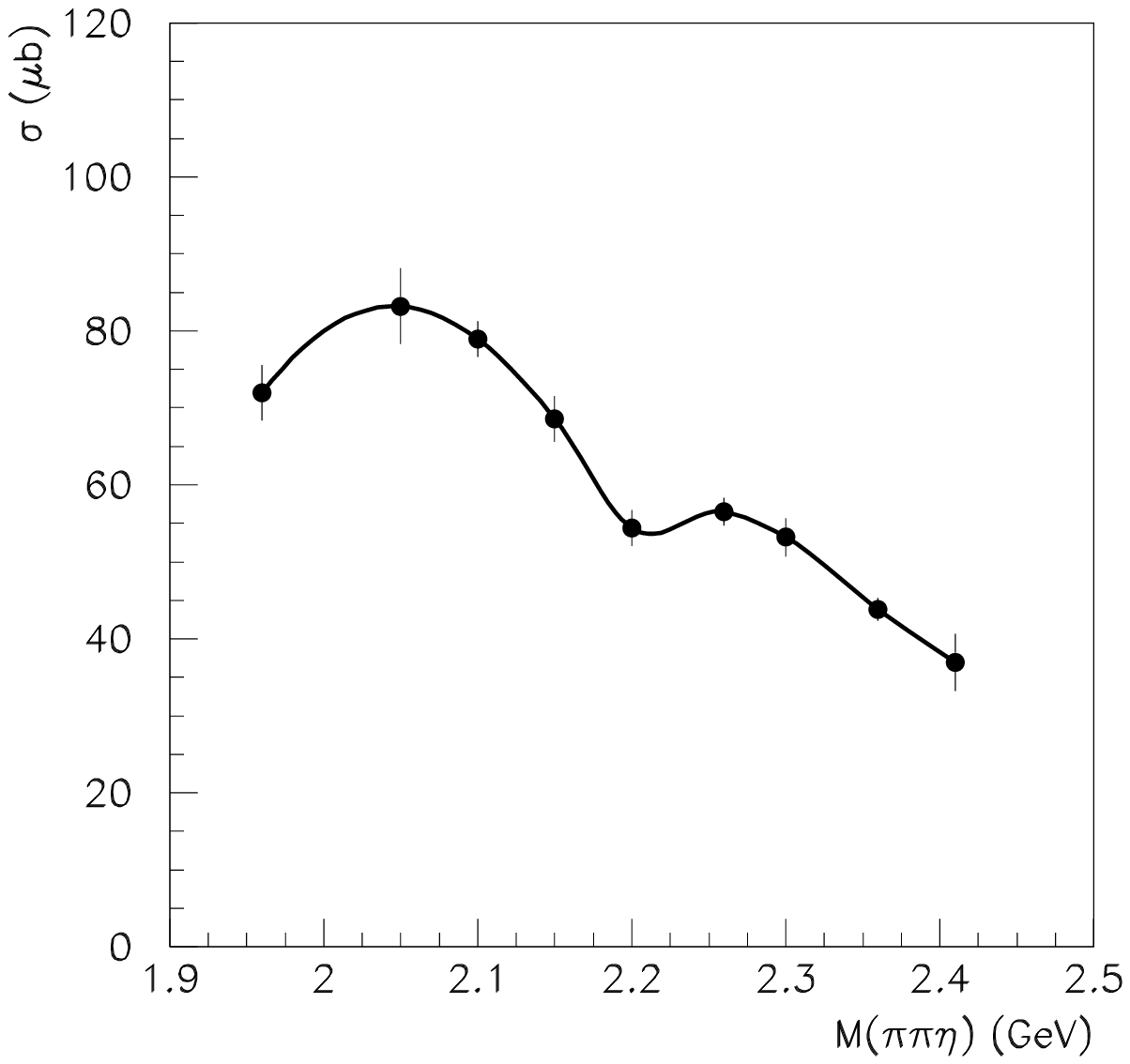}
\vspace*{-1.5cm}
\caption{Cross section for $\bar pp\to\pi^0\pi^0\eta$ with
$\eta\to\gamma\gamma$. }
\end{minipage}
\end{figure}

The only channel which is favourable for hunting $0^{-+}$, $2^{++}$ 
and $2^{-+}$ glueballs and has enough statistics for a full amplitude
partial wave analysis is the $\pi^0\pi^0\eta$ channel. Its cross section is
shown in Fig.4. There are clear enhancements at around 2.05 and 2.3 GeV. 
Note that for a constant amplitude the cross section should decrease steadily
as the energy increases.

For the $\pi^0\pi^0\eta$ channel, 
projections on to $M(\pi\pi)$ and $M(\pi\eta)$ at 900, 1200, 1525 and
1800 MeV/c are shown in Fig.5. The $f_2(1270)$, $a_0(980)$ and $a_2(1320)$
are clearly visible.
The $f_0(980)$ and $f_0(1500)$ are visible on the $M(\pi\pi)$ projection,
but rather weak. As beam momentum increases, the $f_2(1270)$ peak becomes
stronger while the $a_2(1320)$ peak gets weaker; this is a natural
reflection of the rapidly opening phase space for $f_2(1270)\eta$,
whose threshold is at a mass of 1820 MeV.

\begin{figure}[htbp]
\begin{center}\hspace*{-0.cm}
\epsfysize=20cm
\epsffile{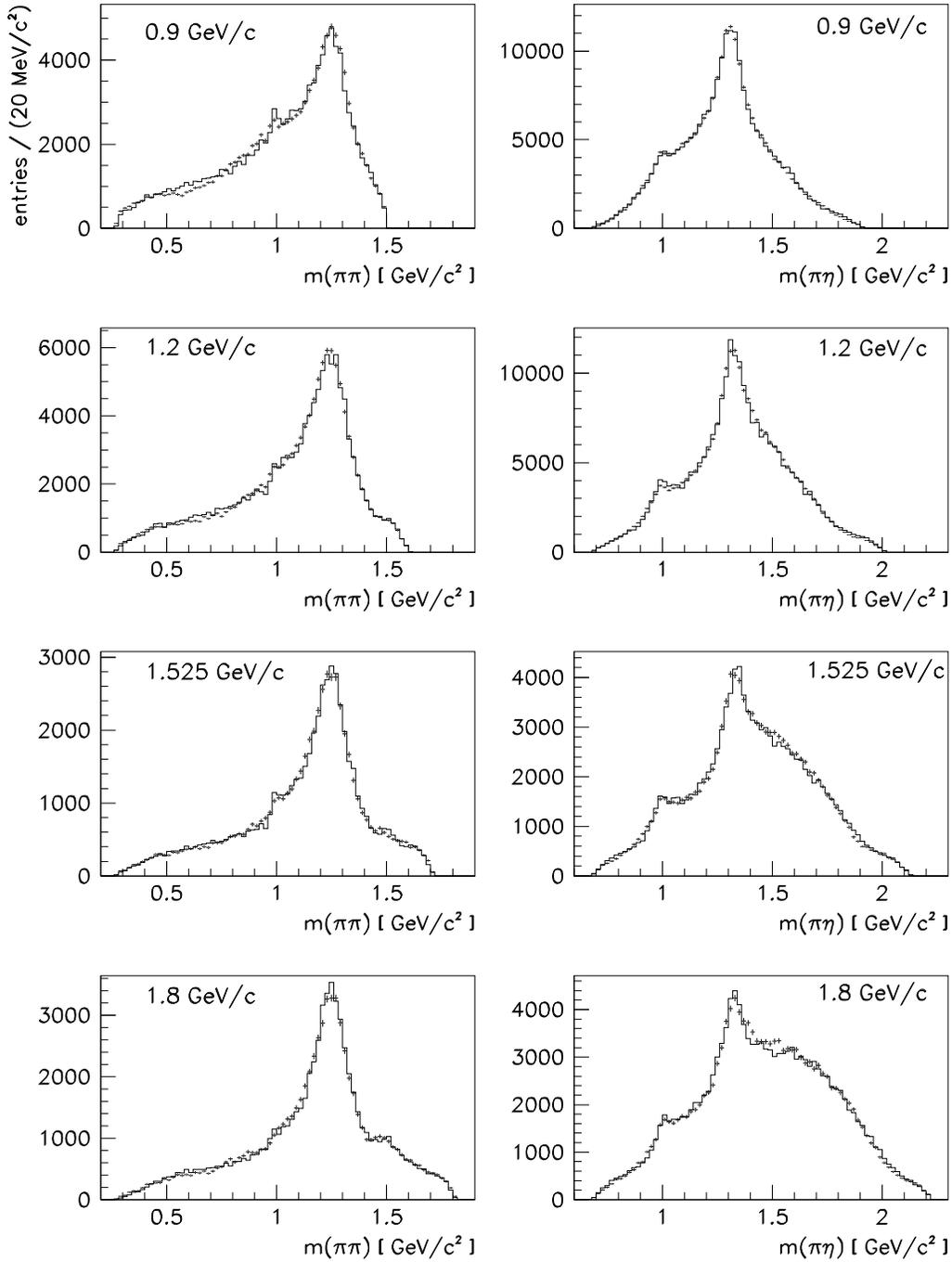}
\end{center}
\caption{Data (points with error bars) and fit (solid line) of invariant
mass spectra for $\pi^0\pi^0$
(1 entry/event) and $\pi^0\eta$ (2 entries/event). Beam momenta are given
in each panel.}  
\end{figure}

Based on these data, a full amplitude analysis describing both production
and decay of these resonances is carried out for each momentum\cite{PPE}.

\section{AMPLITUDE ANALYSIS OF $\bar pp\to\pi^0\pi^0\eta$}

For the $\pi^0\pi^0\eta$ final state, possible $\bar pp$ initial states
are  
$0^{-+}$, $2^{-+}$, $4^{-+}$ etc. for $\bar pp$ spin singlet states,
and $1^{++}$, $2^{++}$, $3^{++}$, $4^{++}$, $5^{++}$ etc. for $\bar pp$
spin triplet states. For our case with center-of-mass energies below 2.41
GeV,
only $0^{-+}$, $2^{-+}$, $1^{++}$, $2^{++}$, $3^{++}$ and $4^{++}$ are
expected to be significant \cite{Hasan} and this has been confirmed in our
analysis;
$4^{-+}$ has been tried, but is not significant.
Their corresponding
$\bar pp$ total angular momentum J, orbital angular momentum L and total
spin angular momentum S in the usual contracted form $^{2S+1}L_J$ are:
$^1S_0$ for $0^{-+}$, $^1D_2$ for $2^{-+}$, $^3P_1$ for $1^{++}$,
$^3P_2$ or $^3F_2$ for $2^{++}$, $^3F_3$ for $3^{++}$, and $^3F_4$ or
$^3H_4$
for $4^{++}$.

Let us choose the reaction rest frame with the z axis along the $\bar p$
beam direction. Then the squared modulus of the total transition amplitude
is the following\cite{Obelix}:
\bea
I&=&|A_{0^{-+}}+A_{2^{-+}}|^2+|A^{M=1}_{1^{++}}+A^{M=1}_{3^{++}}|^2
+|A^{M=-1}_{1^{++}}+A^{M=-1}_{3^{++}}|^2 \nonumber\\
& & +|A^{M=0}_{2^{++}}+A^{M=0}_{4^{++}}|^2
+|A^{M=1}_{2^{++}}+A^{M=1}_{4^{++}}|^2+|A^{M=-1}_{2^{++}}+A^{M=-1}_{4^{++}}|^2
\\\nonumber  
& & +2Re[(A^{M=1}_{2^{++}}+A^{M=1}_{4^{++}})
(A^{M=1}_{1^{++}}+A^{M=1}_{3^{++}})^* 
-(A^{M=-1}_{2^{++}}+A^{M=-1}_{4^{++}})(A^{M=-1}_{1^{++}}+A^{M=-1}_{3^{++}})^*]
\eea
where M is the spin projection on the z-axis.  Partial wave
amplitudes $A_{J^{PC}}$ are constructed from relativistic Lorentz
covariant
tensors, Breit-Wigner functions and Blatt-Weisskopf barrier factors
\cite{Chung,Filip}. The barrier factors use a radius of 1 fm.
The $f_2(1270)\eta$, $a_2(1320)\pi$, $a_0(980)\pi$,
$\sigma\eta$, $f_0(980)\eta$ and $f_0(1500)\eta$ intermediate states are
considered. The $f_0(980)$ is fitted with a Flatt\'e formula using
parameters determined previously \cite{BSZ}. The $\sigma$ is fitted with
the parameterization A of Ref. \cite{BSZ}. Other resonances are fitted
with
simple Breit-Wigner amplitudes  using constant widths.
Full formulae and additional details will be given in Ref.\cite{PPE}.
Based on these formulae, the data at each momentum are fitted by the
maximum likelihood method.

The fit is shown in Fig.5 for the mass spectra for beam momenta at 900,
1200, 1525 and 1800 MeV/c. The quality of the fit for other beam momenta
is similar. The fit to projections is obviously not perfect.
These may be due to additional components such as $a_0(1450)\pi$,
$a_2(1660)\pi$ and even $\hat\rho(1405)\pi$ intermediate states.
An angular momentum decomposition of this small effect is not possible,
but fits including such states produce little effect on the
dominant components.
Since we are mainly interested in scanning the
larger components from $f_2(1270)\eta$, $\sigma\eta$, $a_2\pi$ and
$a_0(980)\pi$ intermediate states, we ignore those smaller contributions
for the present study.

The data points with error bars shown in Fig.6 are our fitted results
for the partial wave cross sections at each momentum
for $\bar pp\to\pi^0\pi^0\eta$ with $\eta\to\gamma\gamma$. Only those  
partial waves with significant contributions are presented.
There are rich peak and dip structures.

\begin{figure}[htbp]
\begin{center}\hspace*{-0.cm}
\epsfysize=19.8cm
\epsffile{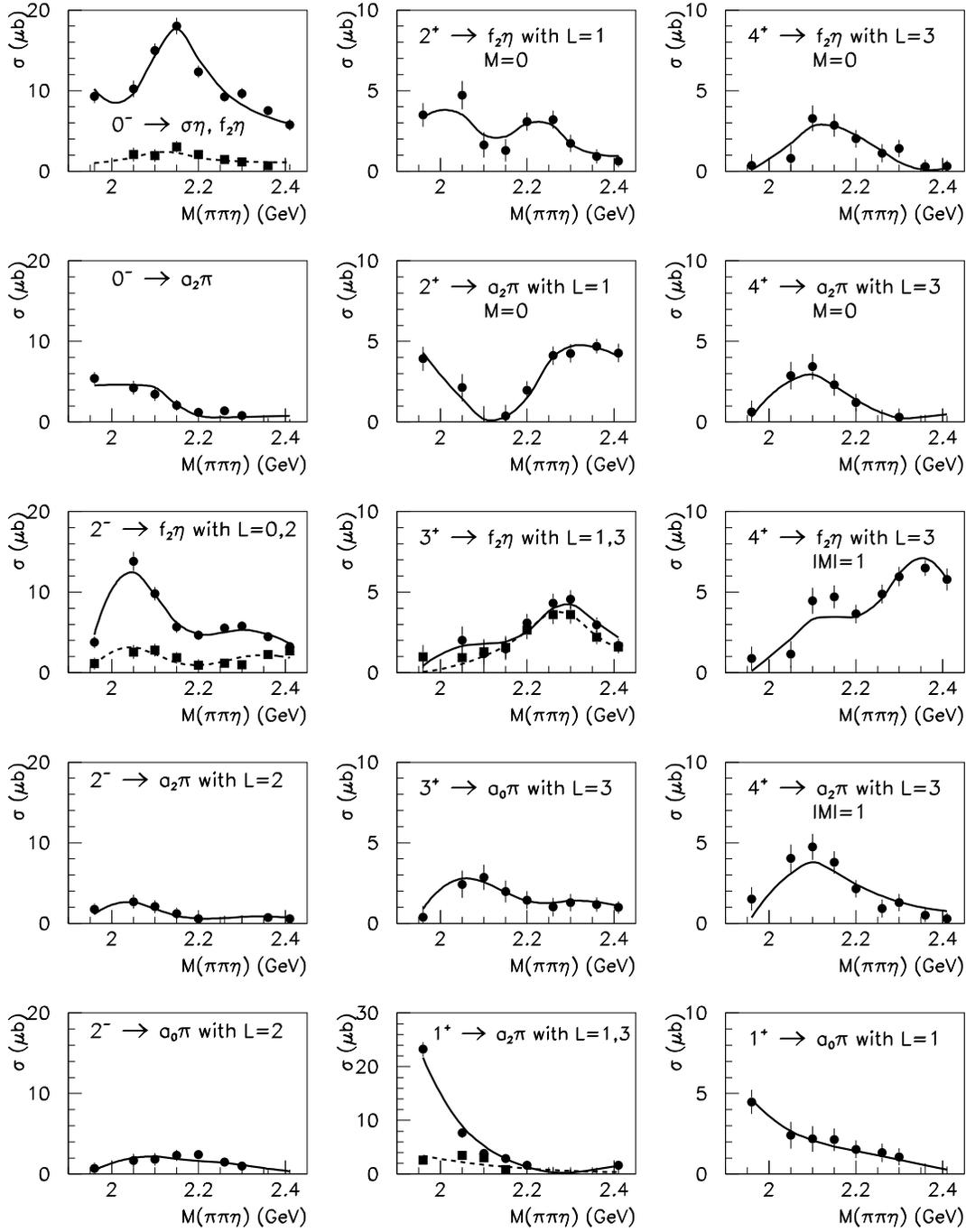}  
\end{center}
\caption{Cross sections for partial waves of significant contributions
to $\bar pp\to\pi^0\pi^0\eta$ with $\eta\to\gamma\gamma$. For diagrams
with two components, the first label (full line) corresponds to the bigger
component. The curves are the fit of Breit-Wigner amplitudes to the data
points in the figure and the relative phases between components.
}
\end{figure}

For $4^{++}$, a peak around 2100 MeV is clear for all $4^{++}$ partial
waves. It is fitted  by a Breit-Wigner amplitude with the mass
and width fixed to the PDG values for the well established $4^+$ resonance
$f_4(2050)$ \cite{PDG}. The shift of the peak position is due
to the centrifugal barrier factors for both initial and final states.
It appears in $f_2\eta$ and $a_2\pi$ with comparable strength.
In addition to the $f_4(2050)$, there is clearly another $4^{++}$ peak 
around $2320$ MeV in $4^+\to f_2\eta$ in the M=1 partial wave.
This may be identified with
the $f_4(2300)$ listed in the Particle Data Tables \cite{PDG}.

A $4^{++}$
resonance around this mass has also been observed be VES in
$\eta\pi^+\pi^-$
in the $\pi A$ reaction \cite{VES}.

For $2^{++}$, two peaks around 2020 MeV and 2350 MeV have masses and
widths
compatible with $f_2(2010)$ and $f_2(2340)$ listed by the PDG \cite{PDG}
as
established particles. In addition, a peak around 2230 MeV shows up    
clearly in the $f_2\eta$ mode. It has a mass compatible with $\xi(2230)$
observed in $J/\Psi$ radiative decays \cite{PDG,BES}, but has a larger
width
of $\sim 150$ MeV.

For $2^{-+}$ and $3^{++}$, both have two peaks around 2050 and 2300 MeV.
No corresponding entries exist for them as yet in the Particle Data
Tables \cite {PDG}.
For $1^{++}$, there is a strong enhancement at the low energy end,
decaying dominantly into $a_2\pi$.

For $0^{-+}$, there seems to be a broad component plus a peak in $\eta
\sigma$
at $\sim 2140$ MeV with width $\sim 150$ MeV.
The broad component may correspond to the
broad $0^{-+}$ object used in describing $J/\Psi$ radiative decays to
$\rho\rho$, $\omega\omega$, $K^*\bar K^*$,
$\phi\phi$ and $\eta\pi\pi$ \cite{BZ}. The peak at 2140 MeV decays
dominantly into $\sigma\eta$.

\begin{table}[hbt]
\caption { Summary of fitted masses, widths and branching ratios
corrected for their unseen decay modes. The mass and width of
$f_4(2050)$ are fixed at PDG values, and the status of the $0^-$ state at
2140 MeV is questionable, as discussed in the text. The $f_1(1700)$ is
beyond the accessible mass range. All states have $I=0$, $G=+1$.}
\begin{center}
\begin{tabular}{ccccccc}
\hline
$J^{PC}$ & M(MeV) & $\Gamma$(MeV) &
${\Gamma_{\bar pp}\Gamma_{f_2\eta}\over\Gamma_{tot}^2} 10^3$ &
${\Gamma_{\bar pp}\Gamma_{a_2\pi}\over\Gamma_{tot}^2} 10^3$ &
${\Gamma_{\bar pp}\Gamma_{\sigma\eta}\over\Gamma_{tot}^2} 10^3$ &
${\Gamma_{\bar pp}\Gamma_{a_0\pi}\over\Gamma_{tot}^2} 10^3$  \\\hline
$4^{++}$ & $2044$ & $208$ & $0.54\pm 0.14$ &
$5.1\pm 0.8$ & - & - \\
$4^{++}$ & $2320\pm 30$ & $220\pm 30$ & $1.3\pm 0.4$ &
$1.0\pm 1.0$ & - & - \\ 
$3^{++}$ & $2000\pm 40$ & $250\pm 40$ & $0.12\pm 0.08$ &
$0.6\pm 0.6$ &  & $0.23\pm 0.11$ \\  
$3^{++}$ & $2280\pm 30$ & $210\pm 30$ & $1.7\pm 0.4$ &
$4.5\pm 2.6$ &  & $0.23\pm 0.19$ \\
$2^{++}$ & $2020\pm 50$ & $200\pm 70$ & $2.1\pm 0.4$ &
$4.3\pm 1.2$ & - & -  \\
$2^{++}$ & $2240\pm 40$ & $170\pm 50$ & $2.5\pm 0.6$ &
$1.6\pm 1.6$ & - & - \\
$2^{++}$ & $2370\pm 50$ & $320\pm 50$ & $0.88\pm 0.64$ &
$16\pm 5$ & - & - \\
$1^{++}$ & $\sim 1700$ & $\sim 270$ &  &  &  & \\
$1^{++}$ & $2340\pm 40$ & $340\pm 40$ & $0.6\pm 0.6$  &
$60\pm 30$ &  & $0.84\pm 0.53$  \\
$0^{-+}$ & $2140\pm 30$ & $150\pm 30$ & $1.9\pm 1.7$ &
$6.0\pm 6.0$ & $10\pm 5$ &  \\   
$2^{-+}$ & $2040\pm 40$ & $190\pm 40$ & $3.0\pm 0.3$ &
$5.0\pm 2.1$ &  & $0.4\pm 0.2$ \\
$2^{-+}$ & $2300\pm 40$ & $270\pm 40$ & $2.8\pm 0.7$ &
$2.0\pm 2.0$ &  & $0.5\pm 0.5$ \\\hline
\end {tabular}
\end{center}
\end{table}

\begin{figure}[htbp]
\begin{center}\hspace*{-0.cm}
\epsfysize=18.5cm
\epsffile{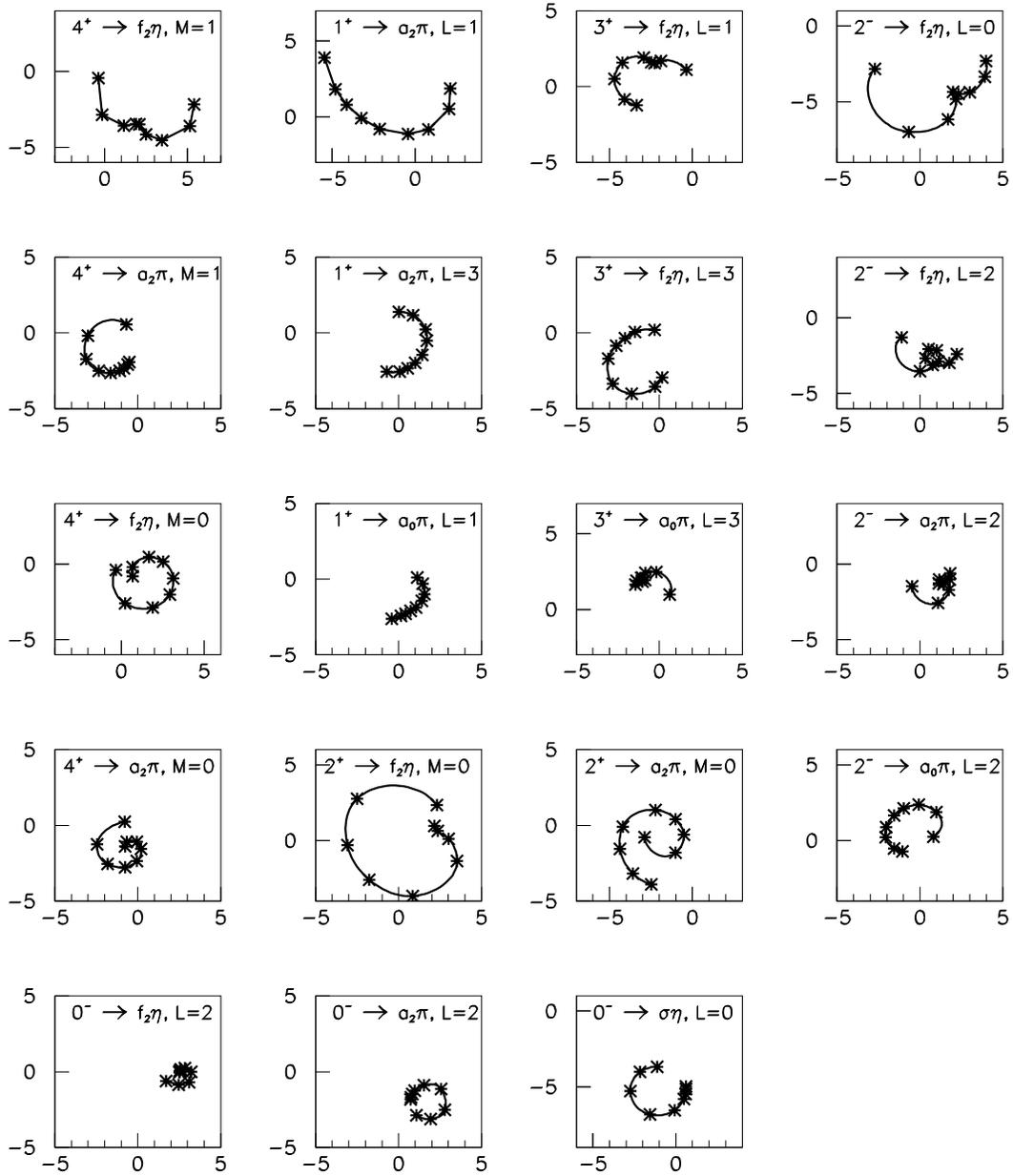}
\end{center}
\caption{Argand plots corresponding to curves of Fig.6}
\end{figure}

Using interfering sums of the Breit-Wigner amplitudes to fit the partial
wave cross sections in Fig.6 as well as relative phases between different
partial waves, we obtain the masses, widths and branching ratios
shown in Table 2. The corresponding Argand plots for
all partial waves are shown in Fig.7. Besides the obvious resonances
mentioned
in previous paragraphs, we need another $1^{++}$ resonance at about 2340
MeV  
with width $\sim 270$ MeV. Without it, we cannot describe the relative
phase
between $1^{++}$ and $4^{++}$ partial waves; also we would need the lower   
$1^{++}$ resonance to be very narrow ($<50$ MeV) in order to explain the
sharp increase  in the $1^{++}$ partial wave cross section at low mass.
In our present fit
with two $1^{++}$ resonances, the $f_1(2340)$ amplitude
interferes destructively with the tail of the lower $1^{++}$ resonance
and causes the sharply decreasing cross section with a broad dip around
2340 MeV. The phase motion caused by this $f_1(2340)$ can be seen clearly
in the Argand plot for the $1^{++} \to a_2\pi$ partial wave in Fig.7.

In Table 2, the branching ratios are calculated at the resonance masses  
and are corrected for their unseen decay modes, except for $a_0(980)$
where $\Gamma_{a_0\pi}=\Gamma_{a_0\pi\to\eta\pi\pi}$.

Among these resonances, the $\eta(2140)$ and $f_2(2240)$ look special.
Both have relative narrow decay width.  The $\eta(2140)$ decays
dominantly into $\eta\sigma$; the $f_2(2240)$ has the largest 
$f_2\eta/a_2\pi$ ratio. These properties suggest that they may
have larger mixing of glue components than other resonances.

\section{ SUMMARY AND OUTLOOK }

In summary, from a full amplitude analysis of $\bar pp\to\pi^0\pi^0\eta$ 
in-flight, we have observed a new decay mode $\eta\pi\pi$ for three
established resonances $f_4(2050)$, $f_2(2010)$ and $f_2(2340)$.
In addition, we have observed 8 new or poorly established resonances
in the energy range from 1960 to 2410 MeV, i.e., $f_4(2320)$, $f_3(2000)$,
$f_3(2280)$, $f_2(2240)$, $f_1(2340)$, $\eta(2140)$, $\eta_2(2040)$
and $\eta_2(2300)$.
Among them, the $0^{-+}$ $\eta(2150)$ has very large decay branching ratio
to $\eta\sigma$; $2^{++}$ $f_2(2230)$ has the largest 
$f_2\eta/a_2\pi$ ratio; both have relative narrow total decay width.
These properties suggest that they may have larger glueball components.

For a further study of the $\eta(2140)$ and $f_2(2240)$, we are going to
reconstruct $\pi^0\pi^0\eta'$ channel from $10\gamma$ events, where we
expect less contamination from other channels. The main purpose here is
to scan the $f_2\eta'$ and $\sigma\eta'$ modes which are also expected to
be favourable decay modes of gluballs.

We are also going to scan $f_2\pi^0$ from $\pi^0\pi^0\pi^0$ final state to
study isovector $q\bar q$ states and scan $f_2\omega$ from
$\pi^0\pi^0\omega$ final state to study isoscalar $q\bar q$ states.
Both $\pi^0\pi^0\pi^0$ and $\pi^0\pi^0\omega$ channels have enough 
statistics for a full amplitude partial wave analysis and have $f_2$
band as their most obvious contribution in their Dalitz plots.

From these analyses of three-body annihilation in-flight, combined with
information from two-body annihilations, we hope to establish the 
$2.0\sim 2.4 GeV/c^2$ meson spectroscopy which is crucial for identifying
gluballs and understanding quark confinement.

\bigskip
{\bf Acknowledgement:} I thank the organizers of the LEAP98 conference for
the invitation to give the talk in such a nice place, Sardinia. I am very
grateful to D.V.Bugg and
A.V.Sarantsev for the fruitful collaboration on data analysis,
and to Crystal Barrel Collaboration for producing the beautiful data.

\end{document}